\documentclass[a4paper,11pt]{article}
\usepackage{aaskaiid}
\usepackage{orcidlink}

\usepackage{amsmath} 
\usepackage{caption}  
\usepackage{multirow}

\usepackage{soul}

\newcommand{\ie}{\it i.e.,}
\newcommand{\eg}{\it e.g.,}

\title{Evolution of AGN Across Cosmic Epochs with the SKAO}
\ShortTitle{Evolution of AGN across cosmic epochs}

\author[1,2]{R. Kondapally\orcidlink{0000-0001-6127-8151}}
\ShortName{Kondapally et al.} 
\author[3]{V. Singh\orcidlink{0000-0002-6040-4993}}
\author[4,5,6]{G. Mazzolari\orcidlink{0009-0005-7383-6655}}
\author[5]{I. Delvecchio\orcidlink{0000-0001-8706-2252}}
\author[7]{A. Datta}
\author[8]{A. Kayal\orcidlink{0000-0001-9851-8243}}
\author[9]{B. Mingo\orcidlink{0000-0001-5649-938X}}
\author[10]{J. Moldon\orcidlink{0000-0002-8079-7608}}
\author[11]{J. Petley}
\author[12]{I. Prandoni\orcidlink{0000-0001-9680-7092}}
\author[13]{K. Rubinur}
\author[14]{S. Shabala}
\author[15]{F. Shankar\orcidlink{0000-0001-8973-5051}}

\affiliation[1]{Centre for Extragalactic Astronomy, Department of Physics, University of Durham, South Road, Durham DH1 3LE, UK}
\affiliation[2]{Institute for Computational Cosmology, Department of Physics, Durham University, South Road, Durham DH1 3LE, UK}
\emailAdd{rohit.kondapally@durham.ac.uk}
\affiliation[3]{Astronomy and Astrophysics Division, Physical Research Laboratory, Ahmedabad, 380009, Gujarat, India}
\emailAdd{veeresh@prl.res.in}
\affiliation[4]{Max-Planck-Institut für extraterrestrische Physik (MPE), Gießenbachstraße 1, 85748 Garching, Germany}
\affiliation[5]{Istituto Nazionale di Astrofisica (INAF) - Osservatorio di Astrofisica e Scienza dello Spazio (OAS), via Gobetti 93/3, I-40129
Bologna, Italy}
\affiliation[6]{Dipartimento di Fisica e Astronomia (DIFA), Università di Bologna, via Gobetti 93/2, I-40129 Bologna, Italy}
\emailAdd{gmazzolari@mpe.mpg.de}
\affiliation[7]{Department of Astronomy Astrophysics and Space Engineering, Indian Institute of Technology Indore, Indore, M.P., India}
\affiliation[8]{Indian Institute of Astrophysics, Block II, Koramangala, Bangalore 560034, India}
\affiliation[9]{Centre for Astrophysics Research, University of Hertfordshire, College Lane, Hatfield, AL10 9AB, UK}
\affiliation[10]{Dpto. Astronomía Extragaláctica, Instituto Astrofísica, Andalucía, Glorieta de la Astronomía s/n, 18008, Granada}
\affiliation[11]{Leiden Observatory, Leiden University, Einsteinweg 55, 2333 CC Leiden, The Netherlands}
\affiliation[12]{INAF-IRA; Via Gobetti 101, 40125 Bologna, Italy}
\affiliation[13]{Institute of Theoretical Astrophysics, University of Oslo, P.O. Box 1029, Blindern, 0315 Oslo, Norway}
\affiliation[14]{School of Natural Sciences, University of Tasmania, Private Bag 37, Hobart, TAS, 7001, Australia}
\affiliation[15]{School of Physics and Astronomy, University of Southampton, Highfield, Southampton, SO17 1BJ, UK}

\abstract{Understanding the evolution of active galactic nuclei (AGN) and their host galaxies across cosmic epochs is one of the key science drivers of extragalactic astronomy. The detection of AGN residing in dusty environments and at high redshifts is difficult due to obscuration and faintness which poses a challenge in understanding AGN evolution across cosmic time. Deep radio continuum surveys (rms noise $<$ 1 $\mu$Jy~beam$^{-1}$) from the Square Kilometre Array Observatory (SKAO) will be an efficient means to detect and study a broad population of AGN across cosmic history. In this chapter, we present radio luminosity functions, source counts, and detection rates of AGN based on the SKAO simulated radio source catalogues. We demonstrate that the SKA-Mid multi-tiered surveys, in particular, reaching sub-$\mu$Jy depths, will allow us to characterise the bulk of the radio-AGN complete down to $L_{\rm{1.4\,GHz}} \sim 10^{23}\,\rm{W\,Hz^{-1}}$ and enable us to probe the evolution of radio-AGN across a wide range of luminosities and all galaxy environments up to $z \sim 6$. Overall, our work highlights the importance of deep multi-tiered SKAO radio continuum surveys for studying the evolution of radio-AGN activity across cosmic time.}

\begin{document}
\maketitle


\section{Introduction}\label{sec:intro}

Active galactic nuclei (AGN) are the manifestation of accretion onto super-massive black holes (SMBHs; $M_{\rm SMBH}$ $\sim$ 10$^{6}$ - 10$^{10}$ $M_{\odot}$) residing at the centres of galaxies and they are thought to play a vital role in the growth of galaxies via feedback mechanisms \citep{Fabian12,Fiore17,Matzeu23}. The well-known correlations between host galaxy parameters such as total stellar mass, bulge mass, central stellar velocity dispersion and the mass of SMBHs \citep{Gebhardt2000,Merritt01,Kormendy13} have been explained by invoking feedback mechanisms wherein AGN can regulate the growth of their host galaxies and play a key role in shaping the observed properties of galaxies in the nearby Universe \citep{McNamara07,Cattaneo2009,Fabian12,Heckman14,Hickox18}. Of particular importance in the lifecycle of massive galaxies 
are `radio-loud AGN' $-$ displaying powerful bi-polar radio jets $-$ feedback from AGN-jet activity is thought to keep massive galaxies quenched. Thus, feedback from AGN is a critical ingredient in galaxy formation models and cosmological simulations \citep[e.g.][]{Bower06,Croton06,Kaviraj2017,Dave2019} that enables us to reproduce the observed local galaxy properties such as galaxy stellar mass function \citep{Bower06,Croton06} and galaxy colours \citep{Cattaneo2006}. 

To study the evolution of AGN across cosmic time it is important to note that the emission from AGN and galaxies can be significantly 
affected by dust extinction in the optical, ultra-violet (UV), and even in the X-rays \citep[e.g.][]{Fabian12,Hickox18}. 
The extinction becomes more prominent in dusty galaxies, providing a biased view of AGN activity; moreover, deep radio surveys have shown an enhanced radio detection fraction for quasars with increasing dust obscuration \citep[e.g.][]{Klindt2019,Fawcett2020,Petley2024,Yue2024}. The mid-infrared colour selection criteria are arguably advantageous in identifying and studying dust-obscured AGN \citep[e.g.][]{Lacy04,Donley12}, however, such studies are typically biased towards high-luminosity AGN. 

Radio continuum observations are unaffected by dust obscuration and also directly trace the synchrotron emission arising from AGN-jets which makes them well-suited to identify and study AGN jet-activity across cosmic time and over a broad range of luminosities. We note that while radio observations offer the only way to identify jetted-AGN across all environments and all epochs, they may miss non-jetted-AGN powered by radiatively-efficient accretion \citep{Padovani2017} emitting mostly thermal radiation from the accretion disk. 
Indeed, no single-band observations alone can provide a complete census of \textit{all} AGN -- multi-wavelength observations are crucial for gaining a more complete understanding.


AGN can broadly be classified into two different modes: (i) radiatively efficient AGN, and (ii) radiatively inefficient AGN, typically based on the Eddington-scaled accretion rate (ratio of the bolometric to the Eddington luminosity; $\lambda_{\rm{Edd}} = L_{\rm{bol}}/L_{\rm{Edd}}$). Radiatively efficient AGN accrete matter at high fractions of the Eddington-scaled accretion rate ($\lambda_{\rm{Edd}} \gtrsim 0.01$), with accretion occurring from a standard geometrically thin, optically thick accretion disk \citep{Shakura1973} and resulting in strong forbidden optical emission lines. A small fraction of radiatively efficient AGN can also display powerful radio jets; traditionally, these sources have been classified as high-excitation radio galaxies (HERGs) based on the nature of their optical spectra \citep{Best2012,Mingo2014}.
Conversely, the radiatively inefficient AGN are often referred to as low-excitation radio galaxies (LERGs) and accrete 
at low rates (typically $\lambda_{\rm{Edd}} \lesssim 0.01$) in an advection dominated accretion flow through a geometrically thick accretion disk \citep{Narayan1995,Yuan2014}. The radiatively inefficient AGN can launch powerful radio jets \citep{Blandford19}, however, as these sources lack a standard stable accretion disk, they may not be easily identified as AGN at other wavebands; radio observations offer a unique window into tracing this population of AGN.  

Over the past two decades, detailed characterisation of the two modes of radio-loud AGN (HERGs and LERGs) has revealed significant differences in their host galaxy properties. LERGs tend to be hosted in massive, quiescent, `red and dead' systems, and are typically found in rich environments such as in galaxy clusters; in contrast, the HERGs tend to be hosted by lower mass, more star-forming systems, and also tend to be more prevalent in poorer environments \citep[e.g.][]{Hardcastle2007,Tasse2008,Smolcic2009,Best2012,Gendre2013,Sabater13,Mingo2014,Ching2017,Williams2015,Williams2018,Croston2019,Magliocchetti22}. However, recent deep observations from SKA pathfinders and precursors have found evidence for more overlap in the accretion and host galaxy properties of radio-loud AGN at higher redshifts than expected (e.g. \citealt{Kondapally2022,Whittam22}; see also discussion in Sect.~\ref{sec:pathfinders}). These results present challenges to our current understanding of radio-AGN and their role in galaxy evolution. Also, present knowledge about radio AGN and their role in influencing galaxy evolution, particularly at higher redshifts, is incomplete.


In this chapter we aim to (i) demonstrate the capability of SKA design baseline Array Assembly 4 (AA4)
to probe the evolution of AGN across cosmic epochs; 
(ii) the role of SKAO multi-frequency radio continuum surveys in unveiling the hidden population of AGN hosted in dust-obscured galaxies;  
(iii) synergies between SKAO and other multi-wavelength facilities in investigating the evolution of AGN across cosmic epochs.
This chapter also offers an update of a chapter on `Exploring AGN Activity over Cosmic Time with the SKA' authored by \cite{Smolcic15} in `Advancing Astrophysics I' book. Moreover, unlike previous work we use a mature set of SKAO specifications and also place an emphasis on the detection of dust-obscured AGN.

\section{Radio continuum surveys: From pathfinders and precursors to the SKAO}\label{sec:pathfinders}
%
Over the past decade or so, SKAO pathfinder and precursor telescopes in both Northern and Southern hemispheres have been carrying out deep as well as wide-area radio continuum surveys ranging from a few square arcminutes to tens of thousands of square degrees across a wide range of frequencies {\it e.g.,} 40-66\,MHz Low Frequency Array (LOFAR) LBA Sky Survey (LoLSS; \citealt{deGasperin2023}), 144\,MHz LOFAR Two-metre Sky Survey (LoTSS; \citealt{Shimwell17,Shimwell22}), 150\,MHz TIFR-GMRT Sky Survey \citep[TGSS;][]{Intema17}, 72$-$231 MHz GaLactic and Extragalactic All-sky MWA survey \citep[GLEAM;][]{Hurley-Walker2017}, 
940\,MHz Evolutionary Map of the Universe \citep[EMU;][]{Norris11}, 1250\,MHz Rapid ASKAP Continuum Survey \citep[RACS;][]{McConnell20}, Jansky Very Large Array (VLA) COSMOS 3.0\,GHz large project \citep{Smolcic17}, 3.0 GHz VLA Sky Survey (VLASS; \citealt{Lacy20}), and 1.3\,GHz MeerKAT International Giga-Hertz Tiered Extragalactic Exploration (MeerKAT; \citealt{Heywood22,Hale25}).
\par 

\begin{figure}
    \centering
    \includegraphics[width=\linewidth]{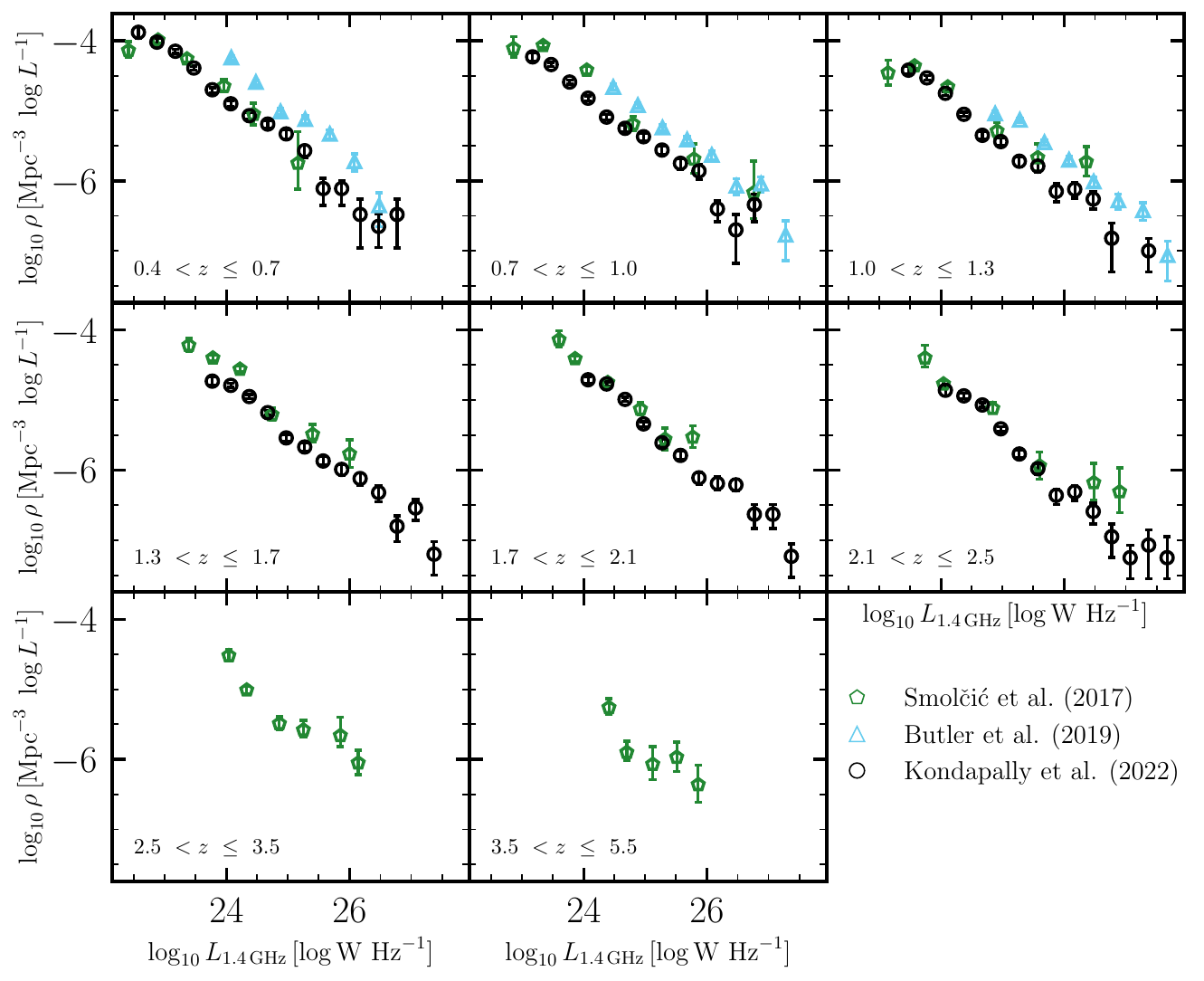}
    \caption{\label{fig:K22_LF_comp}The 1.4\,GHz radio luminosity functions in eight redshift bins over the redshift range $0.4 < z \leq 5.5$. The 150\,MHz radio luminosity functions from \citet{Kondapally2022} across $0.4 < z \leq 2.5$, based on LoTSS Deep Fields, are shown as black circles (scaled to 1.4\,GHz using a spectral index $\alpha = -0.7$). The luminosity functions derived from the VLA-COSMOS 3\,GHz Large program across $0.4 < z \leq 5.5$ by \citet{Smolcic2017_agn} are shown in green. The LFs derived based on ATCA observations by \citet{Butler2019} are shown in cyan. Figure adapted from \citet{Kondapally2022}.}
\end{figure}

Deep radio continuum surveys from SKA pathfinders and precursors over several fields with high-quality multi-wavelength datasets have been crucial in tracing the low luminosity AGN population out to $z \gtrsim 1$. Figure~\ref{fig:K22_LF_comp} shows the evolution of the 1.4\,GHz radio luminosity function (RLF) across the redshift range $0.4 < z \leq 5.5$ from various existing surveys. 
The RLFs (taken from \cite{Kondapally2022}; black open circles) extending out to $z \sim 2.5$ are determined from the deep tier of LoTSS: the LoTSS Deep Fields \citep{Tasse2021,Sabater2021,Kondapally2021,Duncan2021,Best23} that covers $\sim$ 25\,deg$^{2}$ down 
to an rms level of $\sim 20\,\mu$Jy/beam. In Figure~\ref{fig:K22_LF_comp} (depicted via triangles), we also show RLFs 
(from \cite{Butler2019}) compiled using data from 2.1\,GHz observations of the XMM-LSS field by the Australian Telescope Compact Array (ATCA), out to $z \sim 1.4$. The LFs by \citet{Smolcic2017_agn} based on the VLA-COSMOS 3\,GHz Large project are shown in green, providing measurements out to $z \sim 5$. 
We point out that the VLA-COSMOS 3\,GHz Large project and the LoTSS Deep Fields represent the deepest degree-scale radio-continuum surveys that exist. While these surveys have provided the first measurements of the radio luminosity functions for AGN out to $z \sim 5$ (in the case of the VLA-COSMOS project), this is based only on a few hundred sources beyond $z \sim 2.5$ and only on a few tens of sources at $z \gtrsim 4$. Hence, the luminosity function remains poorly constrained at the bright end due to the relatively small areas covered by these deep surveys, and also at the faint end (at high redshift) due to the small numbers of sources.

Deep surveys from LOFAR and MeerKAT have also enabled studies of the different modes of radio-loud AGN (i.e. the LERGs and HERGs) out to $z \sim 2$ and beyond \citep[e.g.][]{Kondapally2022,Mingo2022,Whittam22}. Such studies have revealed that the host galaxies of LERGs and HERGs show considerable overlap at these moderate redshift ($z \sim 2$; \citealt{Kondapally2022,Whittam22,Delvecchio2022}), with a significant population of LERGs hosted by star-forming galaxies rather than quiescent galaxies \citep{Kondapally2022,Kondapally2025}. \citet{Whittam22} also found that the Eddington-scaled accretion rate distributions for LERGs and HERGs showed significant overlap, however, uncertainties in SED fitting of photometry, which was used to derive the classifications, and bolometric luminosities can complicate the interpretation of these results. More recent studies incorporating spectroscopic classifications have shown that the accretion rate distributions appear more similar to studies at low redshift \citep[e.g.][]{Arnaudova2025}. These results highlight that our understanding of the radio-loud AGN population at moderate redshifts may be incomplete based on studies in the local Universe; such studies at earlier epochs have only been possible in recent years. The evolution of the radio-loud population at even higher redshifts towards cosmic dawn remains largely unexplored.

\section{SKAO view of radio-AGN activity across cosmic time}\label{sec:ska_view}
In this chapter, we focus on addressing the open questions highlighted above by discussing the broad population-level studies on the evolution of radio-AGN that will be possible with the SKAO. Specifically, we aim to show that the SKAO will be able to:

\begin{itemize}
\item Chart the complete census of radio-AGN down to $L_{\rm{1.4\,GHz}} \sim 10^{23}\,\rm{W\,Hz^{-1}}$ all the way up to $z \sim 6$ and provide the first robust statistical constraints on their evolution up to $z \sim 6$.
\item Assess the detection and evolution of hitherto, hidden population of AGN in obscured environments across 
large redshifts. 
\item Characterise the evolution of AGN activity at each cosmic epoch as a function of various properties (e.g. stellar mass, star-formation rate, environment) simultaneously.
\end{itemize}

Fulfilling these objectives will allow us to trace in detail how the radio-AGN population evolves across the history of the Universe and to place this in the context of the evolution of the galaxy population to determine the physical processes that govern AGN activity from early epochs to the present day. Achieving these scientific goals requires the combination of a large-area survey that can sample the rarer AGN across a wide range of galaxy environments, and a smaller-area deep survey that can detect and characterise the faint AGN population out to the earliest cosmic epochs. Our scientific objectives can be primarily facilitated by two large-area SKA-Mid (AA4) surveys in band 2: a 1000\,deg$^{2}$ wide-area survey reaching 
to noise-rms of $\sim 1\,\mu$Jy~beam$^{-1}$, and a deep $\sim$ 25\,deg$^{2}$ survey reaching to noise-rms of $\sim 0.2\,\mu$Jy~beam$^{-1}$. These values are based on the proposed SKAO AA4 continuum reference surveys outlined in \cite{Prandoni15} and in the updated version by \citet{Prandoni01.2026.SKA}, which we validate and justify below for our scientific objectives. The sensitivity and angular resolution offered by SKA-Mid AA4 design baseline will be crucial for most of the scientific objectives outlined in this chapter. 
The SKA-Low radio surveys will offer valuable complementary information to AGN studies, although, due to 
lower resolution they will be confusion limited at the sensitivity level of 5$-$10~$\mu$Jy~beam$^{-1}$. 
Nevertheless, SKA-Low wide-area surveys will enable us to perform resolved spectral-index studies for large samples of predominantly brighter and nearby radio-AGN, offering new insights into the physics of AGN jets \citep[see][]{Hardcastle01.2026.SKA}.
To make predictions for the radio-loud AGN population that we can detect in SKAO reference surveys we exploit the Tiered Radio Extragalactic Continuum Simulations (T-RECS; \citealt{Bonaldi19}) considering the aforementioned areas and depths for the wide and deep tiers.

\subsection{Differential source counts}
Radio source counts are commonly used to examine the relative number density of various sub-populations (RL-AGN, RQ-AGN and SFGs) across a range of flux densities and to constrain their evolutionary models \citep{DeZotti10,Condon12}. The source counts extrapolation to fainter flux densities can help us in predicting the radio source populations to be detected in the future SKAO radio continuum surveys.
\begin{figure*}[h]
\includegraphics[angle=0,width=15.0cm,trim={0.0cm 0.0cm 0.0cm 0.0cm},clip]{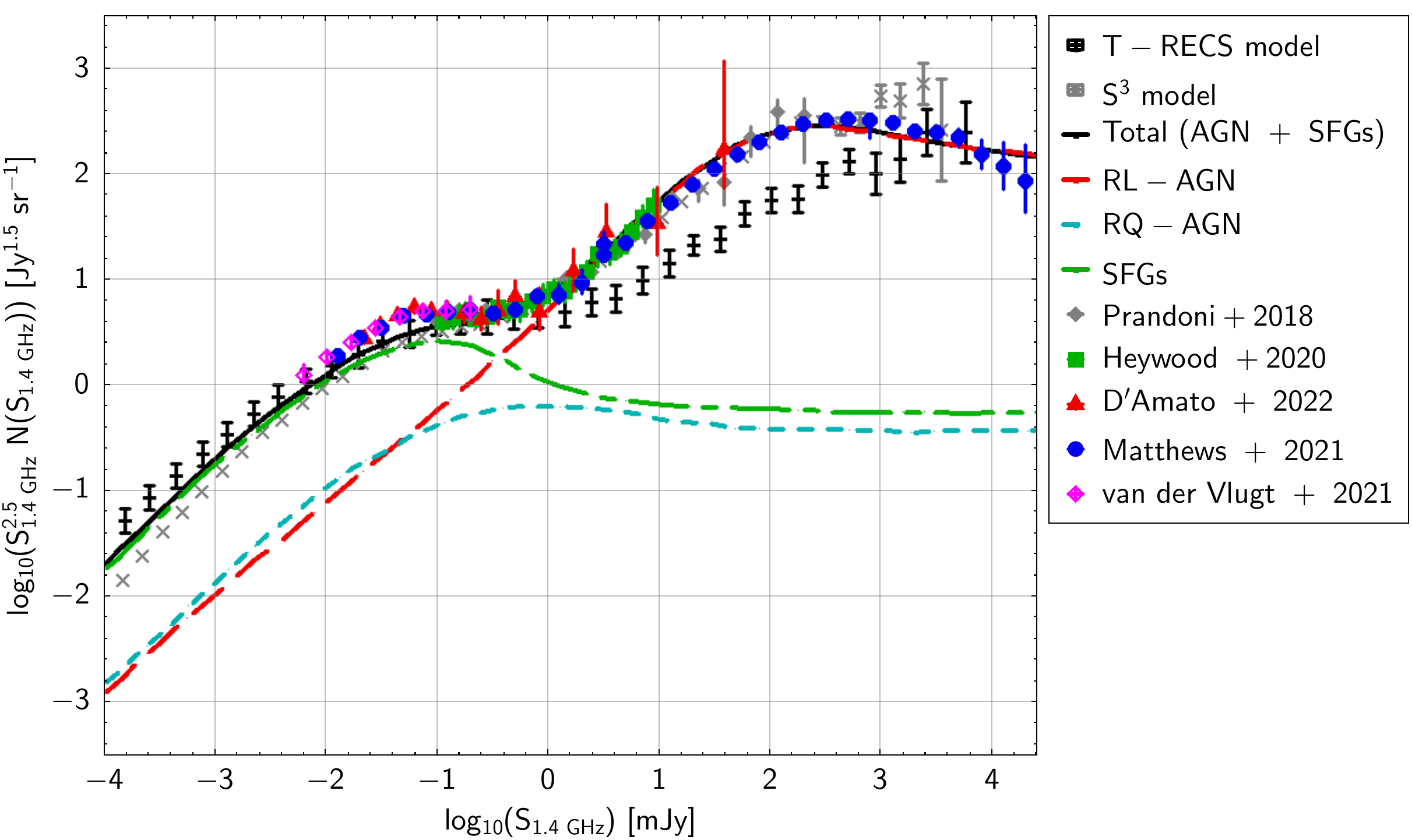}
\caption{The comparison of differential source counts at 1.4 GHz from T-RECS, different radio source sub-populations (RL-AGN, RQ-AGN, SFGs)
models from \cite{Mancuso17}. The available data from recent studies \citep{Prandoni18,Heywood20,Matthews21,vanderVlugt21,DAmato22} are also shown.}
\label{fig:SC}
\end{figure*}
In Figure~\ref{fig:SC} we show 1.4 GHz differential radio source counts derived from some of the recent radio continuum observations in deep fields \citep[e.g.,][]{Prandoni18,Heywood20,vanderVlugt21,Matthews21,DAmato22}.
We also compare these source counts with those derived from SKA simulations {\ie} T-RECS \citep{Bonaldi19} and SKA Simulated Sky \citep[S$^{3}$;][]{Wilman08} as well as semi-empirical model for various sub-populations reported in \cite{Mancuso17}.

It is evident that the source counts at $S_{\rm 1.4~GHz}$ $\geq$ 0.5 mJy is dominated by RL-AGN (see Figure~\ref{fig:SC}).
At bright flux densities, contribution from SFGs and RQ-AGN is nearly insignificant. Although, at fainter flux densities ($S_{\rm 1.4~GHz}$ $<$ 0.5 mJy) contribution from SFGs and RQ-AGN begins to rise.
At the very faint end ($S_{\rm 1.4~GHz}$ $<$ 0.1 mJy), the radio source population is comprised of SFGs
and RQ-AGN, however, the number density of RQ-AGN is nearly one order-of-magnitude lower than that of SFGs.
These results are similar to several previous works \citep[{\it e.g.},][]{Bondi08,Vernstrom16,Smolcic17,Galluzzi25}, although, the relative contribution of SFGs and RQ-AGN is still a matter of debate, and we require equally deep multi-wavelength data to identify RQ-AGN hosted in composite SFGs.
At $S_{\rm 1.4~GHz}$ $>$ 1.0 mJy, T-RECS under-predicts source counts possibly due to its source population inputs derived from the limited sky coverage lacking the complete RL-AGN population \citep[see][]{Bonaldi19}. The S$^{3}$ simulated counts generally match well with the observations, except at fainter flux densities ($S_{\rm 1.4~GHz}$ $<$ 0.01 mJy) where they under-predict the observations. It is worth noting that the deepest available observations from the pathfinders and precursors till date, are limited only to a few ${\mu}$Jy level and the nature of faint radio population is still unexplored.
The SKA-Mid band-2 deep radio continuum surveys offering 0.2 ${\mu}$Jy~beam$^{-1}$ noise-rms and $\sim$0$^{\prime\prime}$.5 angular resolution in AA4 configuration \citep{Prandoni01.2026.SKA},
will open a new window to understand the nature of sub-${\mu}$Jy faint sources without suffering from confusion issues.

\subsection{Evolution of radio-loud AGN: Robust statistical constraints on radio-AGN luminosity function beyond $z > 4$}

\begin{figure}
    \centering
    \includegraphics[width=0.9\textwidth]{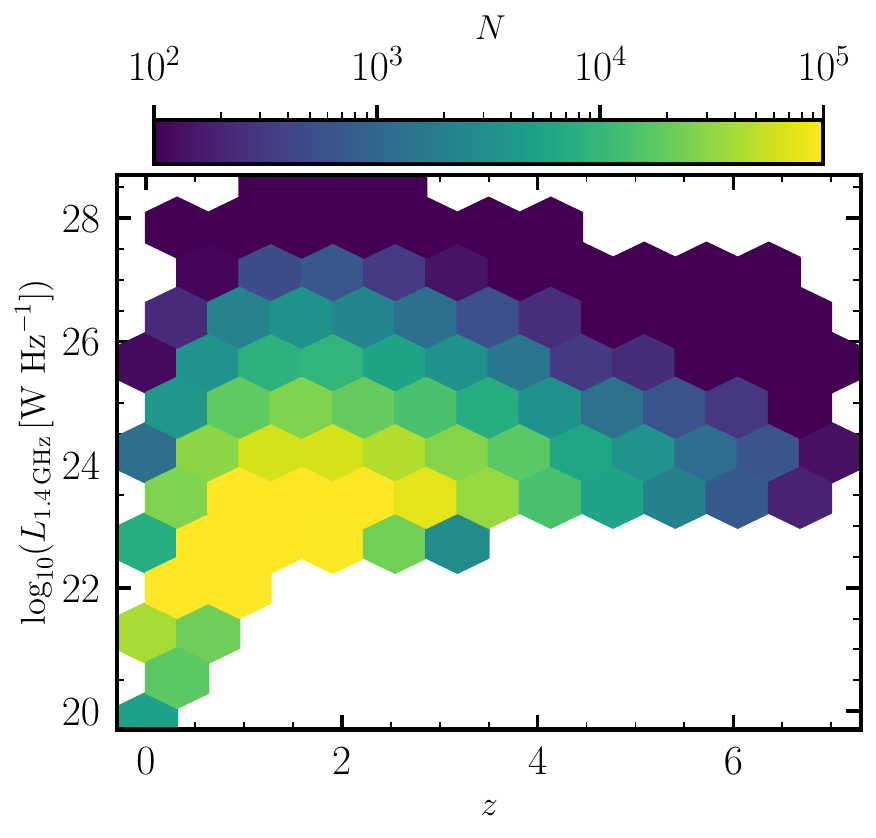}
    \caption{\label{fig:SKA_trecs_L_z}Distribution of sources in the radio luminosity -- redshift plane expected for the chosen SKAO reference radio continuum surveys (using 5$\sigma$ detection; see text). The combination of SKA-Mid band 2   wide and deep surveys will provide a complete census of radio-AGN down to $L_{\rm{1.4\,GHz}} \sim 10^{23}\,\rm{W\,Hz^{-1}}$ all the way up to $z \sim 6$.}
\end{figure}

\begin{figure}
    \centering
    \includegraphics[width=0.8\textwidth]{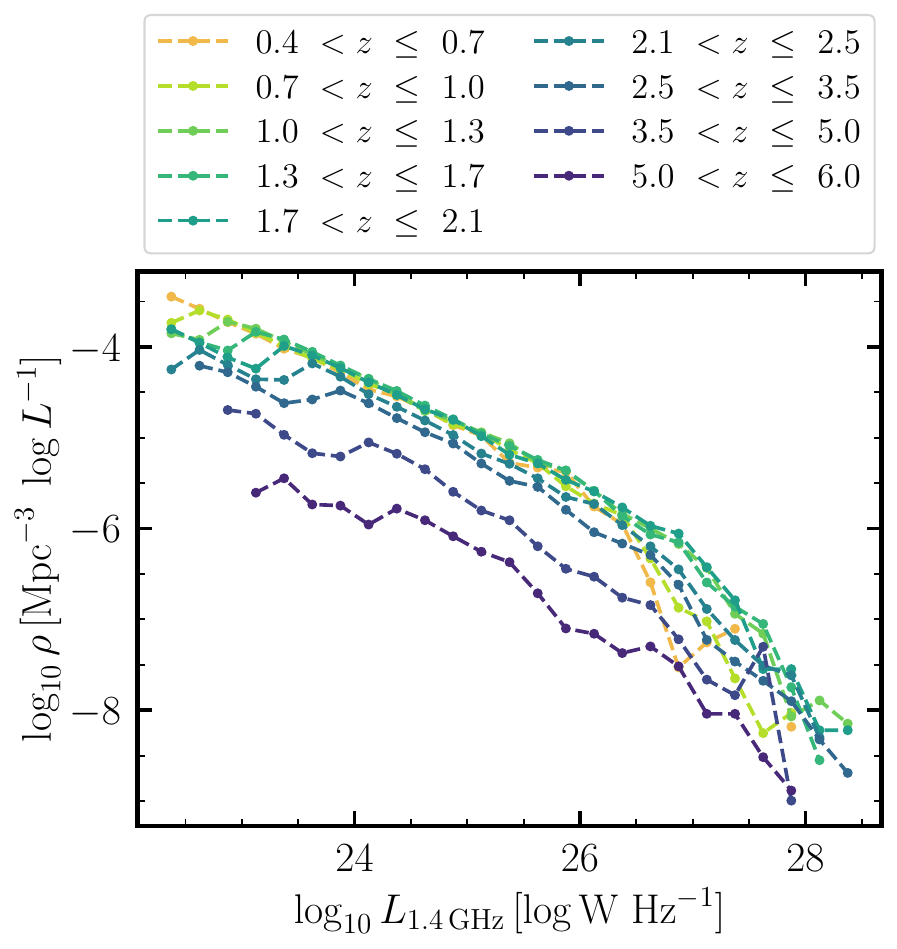}
    \caption{\label{fig:SKA_RLF_full}The cosmic evolution of the 1.4~GHz radio luminosity functions based on 
    the T-RECS predicted source population using the reference SKA-Mid wide and deep tier surveys (see text and \citealt{Prandoni01.2026.SKA}).}
\end{figure}

In Figure~\ref{fig:SKA_trecs_L_z}, we show the binned (as hex-bins) expected distributions of radio-loud AGN in the 1.4\,GHz radio luminosity and redshift plane, based on the assumed survey depths and areas described earlier in this section. The colour bar indicates the number of sources in each bin, where lighter colours indicate a higher density of sources. Using the predicted source population from T-RECS, we construct the expected RLFs for the radio-loud AGN in narrow redshift intervals out to $z \sim 6$, which are shown in Figure~\ref{fig:SKA_RLF_full}. At any given epoch, the luminosity functions 
are expected to reach at least an order of magnitude fainter than those derived from the deepest radio-continuum surveys available to date. We point out that the SKA-Mid band-2 wide-area survey will not only be more sensitive than the exiting deep field radio continuum surveys at similar frequency but also will cover 10-50 times more sky-area, and hence, it will provide more robust constraints at high luminosities and high redshifts, while sampling a wide range of galaxy environments. It is important to note that the shape of the LFs and its detailed evolution are \textit{only indicative} of what we would expect to detect based on the above chosen surveys: these are based on the predictions from the T-RECS simulation which involves extrapolation of the observed source population. We use the LFs in Figure~\ref{fig:SKA_RLF_full} to illustrate the range of parameter space that will be accessible for detailed studies of AGN evolution.

\begin{figure}
    \centering
    \includegraphics[width=0.8\linewidth]{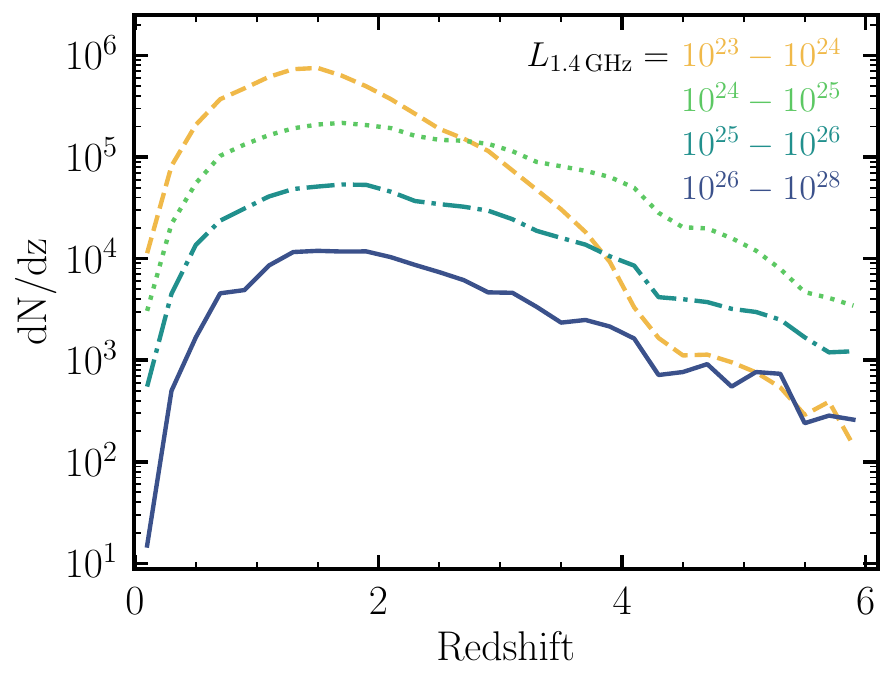}
    \caption{\label{fig:SKA_dndz_lum}Simulated redshift distribution of radio-loud AGN expected to be detected with the SKA-Mid based on the T-RECS simulations and properties of the proposed wide and deep reference continuum surveys noted in the chapter. The expected redshift distributions are shown separately for four radio luminosity bins  across $L_{\rm{1.4\,GHz}} \sim 10^{23}$ -- $10^{28}\,\rm{W\,Hz^{-1}}$. At $z > 5$, we predict that the SKA-Mid observations noted here would detect few hundreds of sources in each luminosity bin; this sample size is sufficient to investigate trends as a function of other galaxy properties (e.g. stellar mass, star-formation rate, etc.).}
\end{figure}


From Figures~\ref{fig:SKA_trecs_L_z} and~\ref{fig:SKA_RLF_full}, it is evident that the SKA-Mid surveys will allow us to detect and study the complete samples of radio-loud AGN down to $L_{\rm{1.4\,GHz}} \sim 10^{23}\,\rm{W\,Hz^{-1}}$ and spanning over five decades in radio luminosity across all redshifts. The large samples of AGN that will be detected, in particular at high redshifts, will enable us to perform detailed characterisation of the radio luminosity function and its shape across a wide range of luminosities beyond $z \gtrsim 4$. 
We predict that the SKA-Mid multi-tiered surveys will be able to detect $> 10^{5}$ AGN across $3 \lesssim z < 5$, and around $10^{4}$ AGN at $5 < z < 6$. For comparison, the latest results from state-of-the-art radio continuum surveys \citep[e.g.][]{Smolcic2017_agn} only contain few tens of sources at $z \gtrsim 5$. Thus, SKA-Mid surveys will provide a fundamental step-forward in our understanding of the evolution of high-redshift radio-loud population by enabling the first robust constraints on the luminosity functions at $z > 5$. 

The radio luminosity function of AGN is also a key ingredient in determining the kinetic luminosity density, which refers to the total kinetic power per unit volume output by the AGN jets across cosmic time \citep[e.g.][]{Shankar2008,Smolcic2009,Best14,Pracy16,Smolcic2017_agn,Kondapally2023}. This is typically determined in the literature by using a scaling relation that converts radio luminosity to jet kinetic power \citep[e.g.][]{Birzan2008,Willott1999,Cavagnolo2010,Godfrey2013}, which is then convolved with the luminosity function and integrated across radio luminosity to determine the volume-averaged kinetic power output. Quantifying this requires both an accurate knowledge of the luminosity function and its evolution, along with an accurate conversion between monochromatic radio luminosity and kinetic power, which are lacking, particularly at higher redshifts. Improved constraints on the latter will be provided by other proposed science cases in this volume which aim to study the physics of AGN jets \citep{Hardcastle01.2026.SKA}, along with accurate size measurements of the radio sources and characterisation of broad-band radio spectra for the AGN (through SKA-Low observations) \citep[e.g.][]{Hardcastle2019,Yates-Jones2023,Turner2023}. For the former, observations discussed in this chapter will robustly characterise the shape and the faint-end of the luminosity function across redshift, which will provide considerably improved constraints on the kinetic heating output by the AGN jets at high redshifts.

\subsection{Map the evolution of radio-AGN activity up to $z \sim 6$}
The SKAO AA4 surveys will also provide large samples of radio-AGN that are complete down to $L_{\rm{1.4\,GHz}} \sim 10^{23}\,\rm{W\,Hz^{-1}}$ at $z \sim 6$ (and even fainter at lower redshifts). This radio luminosity limit is critical as studies at low redshift ($z < 0.5$) have focused on the radio-loud population above this luminosity regime \citep[e.g.][]{Best2005_fagn}. The current deep radio surveys are only able to study complete samples of AGN down to this luminosity limit up to $z \sim 1.5 - 2$, while studies at higher redshifts focus on more luminous AGN. Detailed studies of such faint radio-AGN are not possible from the SKAO pathfinders and precursors surveys. 
The SKA-Mid surveys will provide a transformational increase in our understanding of the faint radio-AGN population out to early epochs. 

The expected redshift distribution of AGN as a function of radio luminosity is shown in Figure~\ref{fig:SKA_dndz_lum}. Across narrow redshift slices, we expect to detect a few hundreds of thousands of AGN at $z \sim 3$ to a few tens of thousands of AGN at $z \sim 5$. With such large samples we can trace in detail how AGN activity depends on multiple relevant properties by binning (e.g. accretion mode, stellar mass, star-formation rate, galaxy age, galaxy environment), simultaneously, across narrow slices of cosmic time to map out the distribution of AGN activity across the galaxy population. At the highest redshifts ($z > 5$), we expect to detect at least a few hundred to a few thousand sources per luminosity bin (see Figure~\ref{fig:SKA_dndz_lum}). This sample size will also be sufficient to investigate trends in radio-AGN activity (e.g. with stellar mass, star-formation rate) by binning as a function of radio luminosity.

%
\par 
At faint flux densities, the so-called `radio-quiet AGN', which do not show clear signs of powerful radio jets, form a significant fraction of the overall AGN population \citep[see][]{Best23}. Whether the radio emission in these sources is dominated by star-formation or by AGN activity is still a topic of debate in the literature \citep[see recent review by][]{Panessa2019}. Recent sub-arcsecond imaging from LOFAR, utilising brightness temperature measurements, has highlighted that the low-frequency radio emission may arise from both star-formation and AGN activity in these sources \citep{Morabito22}. Other VLBI studies at higher frequencies have also found similar results \citep[e.g.][]{HerreraRuiz2017,Radcliffe2018,Deane2024,Saikia2025}. Such brightness-temperature measurements have also shown that even for sources not classified as `radio-loud AGN', a signification fraction of the radio emission may arise from AGN activity \citep{Morabito2025}. Other studies using wide-area radio surveys have also highlighted this, finding considerable overlap between sources that have traditionally been classified as radio-quiet or radio-loud \citep[e.g.][]{Yue2024}. In this chapter, we focus primarily on the radio-loud AGN population, whereas the radio-quiet AGN are discussed in detail by \cite{Panessa01.2026.SKA}.

We emphasise that SKA-Mid band 2 surveys will detect large samples of the (low-luminosity) radio-quiet AGN which will form the bulk of the overall faint AGN population. The sub-arcsecond resolution offered by SKA-Mid band 2 will allow us to disentangle the emission originating from star-formation and AGN processes by resolving the jets on galactic scales. This will also enable the identification of one of the largest clean samples of jetted-AGN. We will be able to trace the prevalence of jetted sources out to the highest redshifts, extending studies at lower redshifts \citep[e.g.][]{Sabater2019,Kondapally2022}. Furthermore, accurate size measurements of radio sources will allow us to trace the evolution of the kinetic luminosity density for the bulk of the jetted sources and quantify the role of jet feedback on galaxy formation at early epochs. We focus primarily on SKA-Mid surveys in this section (and more broadly within the chapter) as the high angular resolution (sub-arcsec level) and deep observations (down to sub-$\mu$Jy level) are necessary to achieve many of the key scientific aims relating to the detection and characterisation of large sample of radio-AGN across cosmic time. SKA-Low observation, with its much poorer angular resolution will be significantly affected by confusion noise.


%
%
%

\subsection{Unveiling AGN in obscured environments}
It is well known that intensely star-forming galaxies possessing large reservoirs of gas and dust can host 
obscured AGN as inflow of gas to the central region triggers AGN activity \citep{Hopkins08,Lacy20}.
The circumnuclear gas and dust causes absorption and scattering of optical, UV and X-ray emission arising 
from the AGN. Thus, optical, UV and X-ray observations pose limitations in unveiling the complete 
AGN population across cosmic epochs. Radio emission being optically thin to gas and dust can allow us to detect 
AGN-jet activity in obscured environments. Using 10 GHz VLA observations \cite{Patil20} detected young AGN-jets of 
a few kpc or smaller in heavily obscured quasars hosted in ultra 
luminous ($L_{\rm bol}$ $\sim$ 10$^{11.7}$ $-$ 10$^{14.2}$ $L_{\odot}$) galaxies residing at 
redshifts ($z$) $\sim$ 0.4 $-$ 3.0. The 0.1-10 GHz radio spectral energy distributions (SEDs) of these 
obscured AGN indicate that the majority of them are Gigahertz-Peaked Spectrum (GPS) sources \citep{Patil22}.
However, most of the radio studies of obscured AGN are limited to radio-bright AGN (S$_{\rm 1.4\,GHz}$ $>$ 1.0 mJy).
Also, shallow wide-area radio surveys have found very low detection rates of dusty galaxies due to 
their limited sensitivities. For instance, using the FIRST survey (5$\sigma$ $\simeq$ 1.0 mJy~beam$^{-1}$), \cite{Gabanyi21} could detect radio emission in only 2.0 per cent of the sources
in their sample of 661 dust-obscured galaxies (DOGs).

Leveraging deep band-3 (250–550 MHz) uGMRT (average noise-rms $\sim$ 30~$\mu$Jy~beam$^{-1}$ and angular resolution
6$^{\prime\prime}$.7 $\times$ 5$^{\prime\prime}$.3) and 1.5 GHz JVLA ((average noise-rms $\sim$ 16~$\mu$Jy~beam$^{-1}$ and angular resolution
4$^{\prime\prime}$.5 $\times$ 4$^{\prime\prime}$.5) radio observations in 2.4 deg$^{2}$ of the XMM-LSS field
\cite{Kayal22} studied radio properties of 321 DOGs selected by $S_{\rm 24~{\mu}m}$/$S_{\rm r-band}$ $\geq$ 1000
criterion \citep{Dey08}.
Both band-3 uGMRT and 1.5 GHz JVLA observations yielded similar detection rates of only
25$-$28 per cent, and demonstrated that, in general, radio emission in DOGs is
faint ($S_{\rm 1.5~GHz}$ $\sim$ 0.035$-$9.33 mJy with a median value of 0.13 mJy for 83/321 radio-detected DOGs)
at sub-mJy level or even lower.
Nevertheless, radio emission in majority of the radio-detected DOGs is inferred to be AGN-jet dominated
considering their radio properties {\eg} high radio luminosities ($L_{\rm 1.5~GHz}$ $>$ 10$^{24}$ W~Hz$^{-1}$),
and  either steep ($\leq$-1.0) or flat ($\geq$-0.5) radio spectral indices (${\alpha}_{\rm 0.4~GHz}^{\rm 1.5~GHz}$).
%
\par
\begin{figure*}[h]
\includegraphics[angle=0,width=8.0cm,trim={0.0cm 0.0cm 0.0cm 0.0cm},clip]{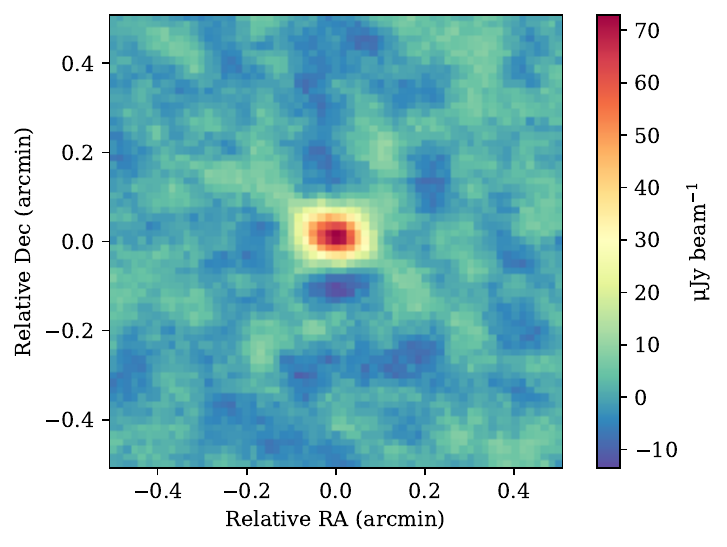}
\includegraphics[angle=0,width=8.0cm,trim={0.0cm 0.0cm 0.0cm 0.0cm},clip]{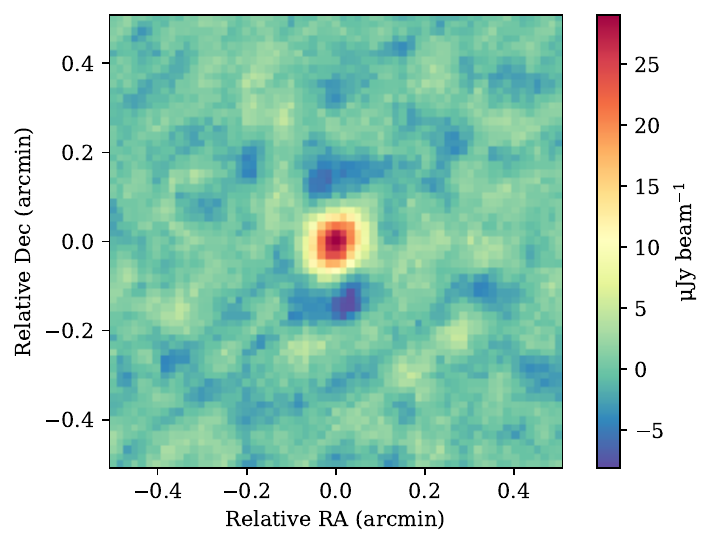}
\caption{The median-stacked images obtained by stacking radio image cutouts centred at the optical
positions of DOGs devoid of radio detection. {\it Left panel}: 400~MHz uGMRT stacked image.
{\it Right panel}: 1.5 GHz JVLA stacked image. This figure is adapted from \cite{Kayal22}.}
\label{fig:stack}
\end{figure*}

We point out that the radio detection rate of DOGs in 1.5 GHz JVLA and band-3 uGMRT is merely 25$-$28 per cent and the radio properties of more than 70 per cent of DOGs remain unexplored. 
For radio-undetected DOGs the median stacked images obtained by stacking 1$^{\prime}$.0 $\times$ 1$^{\prime}$.0 radio
image cutouts at their optical positions show a clear detection of radio emission at SNR $\geq$ 20 with noise-rms
of 3.4 $\mu$Jy beam$^{-1}$ at 400 MHz and 1.5 $\mu$Jy beam$^{-1}$ at 1.5 GHz (see Figure~\ref{fig:stack}), which in turn
implies that, on average, half of the radio-undetected DOGs have 1.5 GHz flux density below 1.5 $\mu$Jy.
%
%
%
More sensitive 1.28 GHz MeerKAT (3.2 ${\mu}$Jy beam$^{-1}$ average noise-rms and 8$^{\prime\prime}$.9 angular resolution)
and band-4 uGMRT (8.4 ${\mu}$Jy~beam$^{-1}$ average noise-rms and 5$^{\prime\prime}$.0 angular resolution) observations performed under the MeerKAT International GHz Tiered Extragalactic Exploration (MIGHTEE) and SuperMIGHTEE projects, respectively, yield nearly 75 and 50 per cent detection rates (Singh et al., {\it in preparation}). However, we note that MIGHTEE and SuperMIGHTEE observations also fail to detect the complete population of DOGs; we require deep SKAO radio continuum observations to obtain a more complete understanding. We also note that the radio emission in a fraction of DOGs may be powered by star-formation, especially sources with relatively lower radio luminosities ($L_{\rm 1.4~GHz}$ $<$ 10$^{24}$~W~Hz$^{-1}$). Indeed, DOGs exhibiting bumps  
(at NIR$-$MIR wavelengths) in their SEDs are known to have intense star-formation \citep{Yutani2022}. Multi-frequency radio observations, that allow us to trace the radio SEDs, would be useful in ascertaining the origin of radio emission as SED modelling can allow us to distinguish 
AGN powered synchrotron emission from star-formation powered by free-free emission \citep{Dey2024}.

\begin{figure*}[h]
\includegraphics[angle=0,width=14.0cm,trim={0.0cm 8.5cm 0.0cm 8.5cm},clip]{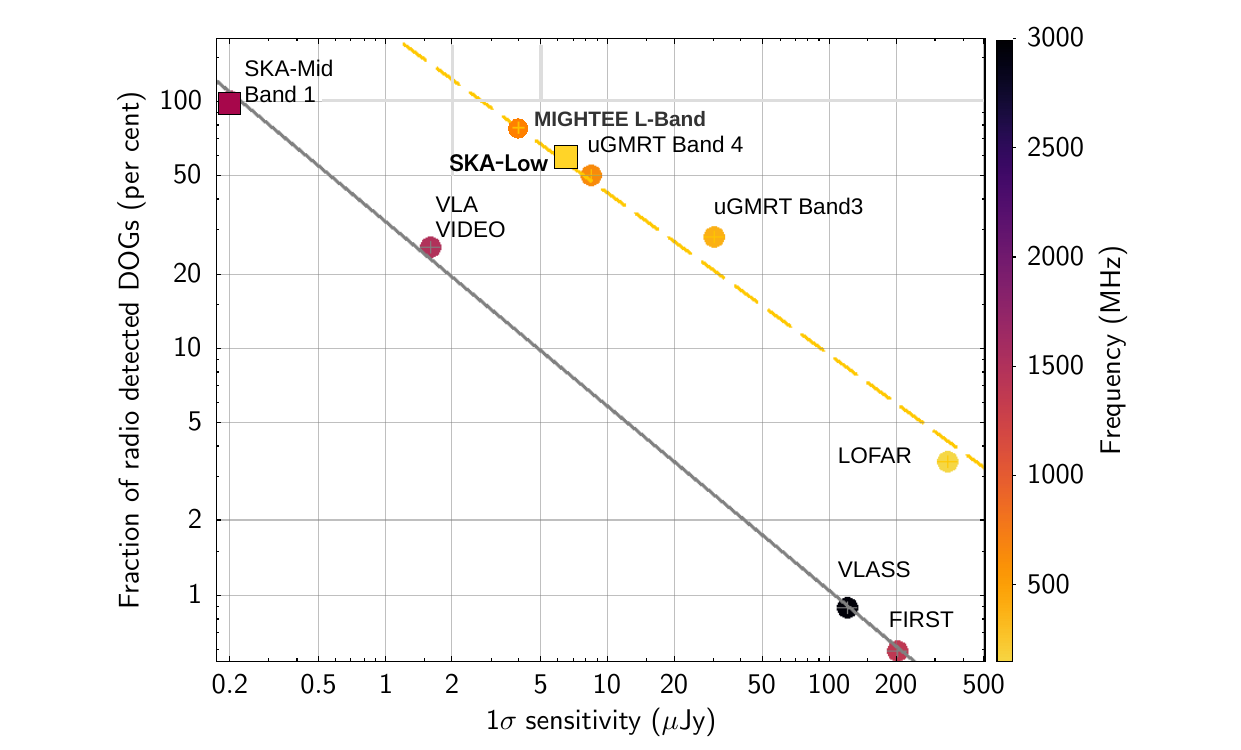}
\caption{The detection rates of dust-obscured galaxies in different radio surveys available in the XMM-LSS field.
The detection rate scales up with sensitivity wherein solid and dotted lines represent the best-fits for high-frequency ($\geq$ 1.4 GHz) and low-frequency ($<$ 1.4 GHz) observations extrapolated to the SKA-Mid and SKA-Low anticipated sensitivities.
The SKA-Mid and SKA-Low data points are depicted by square boxes.}
\label{fig:DetRate}
\end{figure*}

To assess the potential of SKA-Low and SKA-Mid surveys in detecting the radio emission from DOGs  we compare 
the radio-detection rates of 321 DOGs using different radio surveys available in the XMM-LSS field (see Figure~\ref{fig:DetRate}). As expected, the fraction of radio detected DOGs scales up with sensitivity, with both low-frequency as well as high-frequency observations following an increasing trend.
The extrapolation of the trend suggests that the deep SKA-Mid band-2 radio continuum surveys achieving noise-rms 
of 0.2 ${\mu}$Jy~beam$^{-1}$ would detect almost the entire population of DOGs.
We caution that deep SKA-Low observations with integration times of $\gtrsim$ 1 hr in the AA4 configuration 
would result noise-rms of 5-10 $\mu$Jy~beam$^{-1}$ but they would suffer from confusion limits.
Therefore, deep SKA-Low observations are unlikely to achieve 100 per cent detection
rate but they can be advantageous in detecting steep spectrum radio sources especially at higher redshifts.
Furthermore, combination of SKA-Low and SKA-Mid observations would allow us to study radio SEDs of DOGs that are 
essential to ascertain the origin of radio emission {\ie} AGN versus star-formation.
\subsection{Obscured AGN at high-redshifts}
Deep X-ray surveys have shown that the fraction of heavily obscured AGN with equivalent hydrogen column density ($N_{H}) > 10^{23}\rm{cm}^{-2}$) and high X-ray luminosity $L_X>10^{44}$ erg/s (i.e. bolometric luminosity $L_{\rm bol}>10^{45}$ erg/s) increases from $10-20\%$ in the local Universe to $80-90\%$ at $z\sim4$ \citep{Vito18}. 
The analytical studies and numerical simulations \citep{Gilli22,Ni20} addressing the evolution of the physical properties of the inter-stellar medium (ISM) around AGN also support this finding; the latter further predicts that the fraction of obscured AGN may exceed 99\% at $z>6$. Recent works, exploiting \textit{James Webb Space Telescope} (JWST) spectroscopic and photometric data, have found that the obscured AGN number density at $z>5$ is even larger than previous estimates derived from X-ray studies \citep{Akins24,Lyu23,yang23}, and this is more in line with the results from simulations and models \citep{Sijacki15, shankar14, Volonteri16}. All these results potentially suggest that there exists a population of heavily obscured AGN such that even the deepest X-ray observations would miss them. 

The advantage of radio emission being almost unaffected by obscuration is that we can detect AGN activity even in the most obscured environments.
The results from analytical models suggest that the deep continuum radio surveys can unveil a large number of obscured AGN, even those missed by other selection techniques. In \cite{Mazzolari24}, 
by taking the radio and X-ray images of some known extragalactic deep fields, they showed that, on average, the surface density of Compton-thick (CTK) AGN detectable in the existing radio images is $\sim 10$ times higher than that in the corresponding X-ray images. While X-ray selection is generally more effective in detecting unobscured, radiatively efficient AGN, radio emission can more easily unveil the most obscured sources, even those for which X-rays are completely absorbed \citep[like CTK AGN, see the results in][]{Mazzolari26_ctk}. Additionally, while the AGN radio selection might be biased towards low accretion rates \citep{Kondapally2022,Best23}, the combination of radio-excess selection and multi-wavelength datasets have demonstrated the ability to also select radio-quiet and radiatively efficient AGN (see Sect.~\ref{sec:radiosel}).

In this sense, the observations that SKA will perform will be transformative. \cite{Mazzolari24} simulated the number of AGN and CTK-AGN that will be detectable with the SKAO considering three different tiers -- 0.05$\rm \mu Jy$ noise-rms over 1.0 $\rm deg^2$ (ultra-deep survey), 0.2$\rm \mu Jy$ noise-rms over 10-30 $\rm deg^2$ (deep survey), 1$\rm \mu Jy$ noise-rms over $\rm 10^3 \ deg^2$ (wide survey), respectively \citep{Prandoni15,Prandoni01.2026.SKA}. 
They made predictions for three different redshift ranges: $z>3$, $z>6$, and $z>10$. The results for the three tiers are reported in Table~\ref{tab:SKA_pred_compact}. All the predictions are computed assuming a flat sensitivity over the survey area. With the wide SKA survey thousands of AGN are expected to be detected at $z>6$ and a significant fraction of them are expected to be CTK AGN (at least $>45\%$). With such SKAO surveys, it will be possible to trace the evolution of the obscured AGN population from the end of the reionization through the cosmic noon, down to the present universe, and combine the findings obtained in the radio band with those obtained using obscured AGN selections in the X-rays or mid-infrared. In particular, the $z>6$ universe is still very unconstrained, particularly in the X-ray and radio waves.
The possibility to identify hundreds or thousands of AGN and CTK AGN at $z>6$ will transform our understanding of the co-evolution of SMBHs with their galaxies in the first Gyr of cosmic time. 
Also at $z>10$, a redshift range that was almost completely unexplored before the advent of JWST, the SKAO is expected to be able to detect several tens of radio-quiet AGN, opening a completely new window.

\begin{table}[h!]
\centering
\small 
\setlength{\tabcolsep}{2pt} 
\begin{tabular}{ccccccccccccc}
\multicolumn{13}{c}{\textbf{The SKA-Mid predictions for high-$z$ obscured AGN}} \\
\hline
Area (deg$^2$) & Sens ($5\sigma$) [$\mu$Jy] & \multicolumn{3}{c}{$z>3$} &  & \multicolumn{3}{c}{$z>6$} & & \multicolumn{3}{c}{$z>10$} \\
\cline{3-5} \cline{7-9} \cline{11-13}
& & AGN & CTK & $\log L_{\rm bol}$ & & AGN & CTK & $\log L_{\rm bol}$ && AGN & CTK & $\log L_{\rm bol}$ \\
\hline
1    & 0.25 & 1870 & 831   & 43.8 && 34  & 15 & 44.4 && 2  & 1 & 44.8 \\
20   & 1.0  & 13780 & 6120  & 44.4 && 220 & 98 & 44.9 && 6  & 3 & 45.6 \\
1000 & 5.0  & 182000 & 81000 & 45.1 && 1980 & 880 & 45.4 && 35 & 16 & 46.6 \\
\hline
\end{tabular}
\caption{\footnotesize
The expected number of high-$z$ AGN in three 1.4~GHz continuum surveys with SKA-Mid band-2 (as detailed in the text).
The areas and sensitivities are in the units of deg$^2$ and $\mu$Jy, respectively.
$\log L_{\rm bol}$ gives the median bolometric luminosity of detected AGN (erg\,s$^{-1}$).
}
\label{tab:SKA_pred_compact}
\end{table}

\section{Selection of radio-AGN with SKAO and other multi-wavelength observatories}\label{sec:radiosel}

The results presented in the previous sections can only be obtained if AGN can be be properly identified among the many new radio sources that the SKAO will detect. The combination of radio data from the SKAO with multi-wavelength data from current and next-generation facilities will be crucial for selecting distant AGN and distinguishing them from normal SFGs, the main contaminants, particularly at faint radio fluxes. SKAO surveys are planned to overlap with forthcoming wide-area, sensitive optical and near-infrared surveys, such as those by the Vera C. Rubin Observatory Legacy Survey of Space and Time (LSST) and Euclid, but also with the deep surveys performed by JWST NIRCam and MIRI. 
The LSST will perform photometric imaging of entire southern hemisphere sky repeatedly in six filters ($u$, $g$, $r$, $i$, $z$, and $y$; covering 320$-$1060 nm) with time-lapse of few nights and detecting variable sources including AGN. In addition to the wide survey, LSST will also perform deep imaging in five deep-drilling fields (DDFs) namely COSMOS, W-CDFS, ELAIS-S1, XMM-LSS and EDF-S covering a total of nearly 10$-$20 deg$^{2}$ sky area \citep{Brandt18}.  
Euclid is performing optical and near-infrared imaging as well as NIR spectroscopy over the sky-area of nearly 14000 deg$^{2}$ (the wide survey) and 50 deg$^{2}$ of deep fields \citep{Mellier25}. 
The unprecedented sensitivity of both the LSST Wide and in Deep Drilling Field surveys ($r$ band mag $\sim$ 27$-$28) will enable us to detect the high-$z$ AGN population including those residing in obscure environments that will in turn constrain the AGN luminosity function.

The possibility to constrain both the radio and the rest frame optical and near-infrared emission of the sources detected with the SKAO will enable the application of radio-excess (REX) selection techniques that are able to distinguish between accretion, and SF-dominated radio sources \citep{bonzini13,Smolcic17,Delvecchio21,Best23,zhu23,Zhang25,Eberhard25, Mazzolari26_ctk}. These radio-excess selections are able to reliably identify not only the most powerful radio AGN but also a consistent fraction of the RQ-AGN population provided that they possess a dominant AGN-driven radio component, independently of which exact mechanism is responsible for it \citep{Whittam22,Lyu22,Zhang25}. In addition, the development of new and refined SED fitting codes and techniques will also be crucial for identifying AGN among the sources that the SKAO will detect in its radio-continuum surveys (provided that imaging in other wavelengths is also available). So far, the AGN radio emission has been implemented only in some SED-fitting codes, such as \texttt{CIGALE} \citep{boquien19}. A fully consistent modelling of the AGN radio emission in SED-fitting codes is hard, given that, if jet-driven, radio emission is a function of jet power, jet orientation, source age, and environment, none of which can be extracted from the multi-band SED. However, it would be ideal to implement in SED-fitting codes a modelling of the radio emission that could, for example, directly provide an estimate of the AGN-related radio excess, taking into account the constraints on the SF radio emission obtained from the multi-band fit. As demonstrated in this Science Book, the depth of SKAO observations and the possibility of having deep radio images at different frequencies will be extremely useful for constraining the presence of an active SMBH in galaxies that do not show AGN signatures at other wavelengths (for example due to obscuration effects).

Additionally, the combination of mm ALMA and NOEMA follow-up observations, or far-infrared PRIMA surveys \citep{Moullet2023}, with the SKAO will verify whether the far-infrared-radio correlation, typically used to distinguish between AGNs and SFGs, remains valid at high redshifts, as well as for low stellar masses \citep[$M_{*}<10^9\ \rm M_{\odot}$ ][]{Delvecchio21} and SFRs. As an example, the sensitivity of the SKAO deep surveys will allow detection of galaxies with SFR $\rm \sim 10 - 100 M_{\odot}$/yr in galaxies up to $z\sim 3-4$, and SFR $\rm \geq100 M_{\odot}$/yr at $z\sim 6-7$ \citep{mcalpine15,Prandoni01.2026.SKA}.

JWST and ELT observations will also enable spectroscopic follow-up of radio-selected, high-z AGN candidates to determine their redshifts and ultimately measure the radio AGN luminosity function at $z>6$. Furthermore, thanks to SKAO, we expect to reveal the radio emission of JWST-discovered AGN, which so far appear to be elusive at X-ray and radio wavelengths \citep[e.g.][]{Maiolino2025,Lambrides2024}. Indeed, the population of high-z AGN selected using JWST photometry and spectroscopy, including also the so-called Little Red Dots (LRDs), seems to be characterized by a different SED compared to standard AGN, with a peculiar weakness in the X-rays, UV, but also in the radio band \citep[e.g.][]{Matthee2024,Kocevski2025}. However, it is still not clear whether this is due to intrinsically different accretion properties (such as super Eddington accretion) or due to the presence of an obscuring dense cocoon of gas close to the central SMBH. Current analysis did not reveal, even with stacking techniques on the deepest radio images available, and reaching the $\sim$100 nJy noise level, any radio emission in these sources \citep{Mazzolari25, Gloudemans25,Perger24}. Instead, ultra-deep SKAO surveys may help detect their radio emission and reveal the origin of this apparent weakness, which is discussed in further detail in a dedicated chapter within this volume by \citet{Mazzolari01.2026.SKA}.

The combination of the SKAO radio observations with present and future wide and deep X-ray surveys from eRosita \citep{Predehl21}, New-Athena \citep{Cruise25}, and AXIS \citep{Reynolds24} will enable improved AGN identification. In particular, the availability of X-ray observations will enable a precise measurement of the obscuring column density of the sources, and therefore facilitate the study of the evolution of obscured AGNs. The combination of both X-ray and radio observations will enable a detailed study of the processes associated with the feedback mechanisms, such as the formation of X-ray cavities resulting from the launch of radio jets \citep{Birzan2020,Ubertosi2025}.

In addition to multi-wavelength surveys, large spectroscopic surveys, such as those with the SDSS, have been vital in characterising redshifts and properties of radio-detected AGN and their host galaxies \citep[e.g.][]{Best2012,Pracy16}. In the coming years, \textit{targetted} spectroscopic surveys of radio-detected sources from LOFAR (with WEAVE-LOFAR; \citealt{Smith2016}) and MeerKAT (with ORCHIDSS; \citealt{Duncan2023}) will provide large spectroscopic samples of radio-AGN. Similar large-area and targetted spectroscopic surveys in the SKAO era will be crucial in characterising the AGN population detected by SKAO surveys in detail; this is discussed further by \citet{Duncan01.2026.SKA}.

In the absence of other multi-band AGN diagnostics and observations, the identification of AGN and CTK AGN using SKAO data may rely on either multi-frequency radio information, or high-resolution (SKA 10~GHz observations will provide angular resolution of 0.05$^{\prime\prime}$-0.1$^{\prime\prime}$ compared to the 0.4$^{\prime\prime}$-0.5$^{\prime\prime}$ of 1.4~GHz observations) and/or multi-epoch follow-ups. All these diagnostics are amply used in the literature. Flat or convex radio spectral indices would point to the presence of compact AGN cores \citep{odea21}. High-resolution follow-ups can pinpoint AGN through the measurement of brightness temperatures \citep[see][]{Morabito22, Morabito2025,Radcliffe2018,HerreraRuiz2017,Saikia2025,Peluso2025}, while multi-epoch observations may identify those variable AGN at $\mu$Jy flux density levels \citep{radcliffe19b}. In particular the combination of EVN and SKA-AA4 will allow the measurement of brightness temperatures $T_{b}>10^5$ K in $L_{1.4GHz}>8\times 10^{24}\ \rm W/Hz$ (or equivalently $L_{1.4GHz}>10^{41}\ \rm erg/s$) and $z\sim6$ QSO with only few tens of minutes of observations \citep[see][]{Spingola01.2026.SKA}.

\section{Conclusions}

The capabilities offered by the SKAO surveys will enable us to perform population-based studies of radio-AGN and their evolution across cosmic time. In this chapter, we have highlighted the need for multi-tiered SKAO radio continuum surveys (focusing mainly on the SKA-Mid): 1000 deg$^{2}$ wide-area survey down to $\sigma \sim 1\,\mu$Jy/beam and a $\sim 25$\,deg$^{2}$ deep survey down to $\sigma \sim 0.2\,\mu$Jy/beam (both at 1.4\,GHz), to transform our view of the faint raido-AGN population at the earliest cosmic epochs. Using T-RECS simulated radio source catalogues, we have shown that such a tiered survey will provide a complete census of the radio-loud AGN population down to $L_{\rm 1.4~GHz}$ $\sim$ 10$^{23}$ W~Hz$^{-1}$ all the way up to $z \sim 6$. The large samples of AGN detected with the SKA-Mid surveys will: (i) provide the tightest constraints on the evolution of the radio luminosity function up to $z \sim 6$; (ii) enable robust constraints on the faint-end slope of the luminosity function which will be vital for quantifying the kinetic feedback across redshift; (iii) provide the largest clean sample of jetted-AGN, including those within SFGs; and (iv) map the distribution of radio-AGN activity as a function of various galaxy properties (e.g.stellar mass, star-formation rate, galaxy environment) in unprecedented detail across all epochs. We have also emphasised that the SKAO will offer new insights into AGN population residing in obscured environments and argued that, unlike, existing radio continuum surveys from the pathfinders and precursors limited up to a few $\mu$Jy, SKAO continuum surveys that can achieve sub-$\mu$Jy depths will provide a more complete census of obscured AGN up to $z \sim 6$. Synergies with current and planned multi-wavelength surveys will be crucial in achieving many of the key scientific objectives in studies of the evolution of radio-AGN.

\section*{Acknowledgements}
RK acknowledges support from the Leverhulme Trust through a Leverhulme Early Career Fellowship. ID acknowledges funding by the European Union – NextGenerationEU, RRF M4C2 1.1, Project 2022JZJBHM: "AGN-sCAN: zooming-in on the AGN-galaxy connection since the cosmic noon" - CUP C53D23001120006. BM acknowledges support from UKRI STFC for an Ernest Rutherford Fellowship [grant number ST/Z510257/1]. VS acknowledges that the research work at the Physical Research Laboratory, Ahmedabad, is funded by the Department of Space, Government of India. GM acknowledges funding by the European Union (ERC APEX, 101164796). JM acknowledges financial support from the Severo Ochoa grant CEX2021-001131-S and from grant PID2023-147883NB-C21, funded by MCIU/AEI/ 10.13039/501100011033, as well as support through ERDF/EU.

\bibliographystyle{abbrvnat-maxbibnames4}
\bibliography{chapter} 

\end{document}